\documentclass{PoS}
\usepackage{cite}

\newcommand{\pion}{\pi^0}
\newcommand{\process}{p^{\uparrow}p\to p\pion X}
\newcommand{\esum}{E_{sum}}
\newcommand{\xp}[1]{#1_{p}}
\newcommand{\xh}[1]{#1_{\pi}}
\newcommand{\phih}{\xh{\phi}}
\newcommand{\phip}{\xp{\phi}}
\newcommand{\delphi}{\Delta\phi}
\newcommand{\asym}{A_{p\pi}}
\newcommand{\modulation}{\cos\phip\cos\delphi}

\title{
Transverse Spin Asymmetries in the $\process$ Process at STAR
}

\ShortTitle{$\process$ Transverse Spin Asymmetries at STAR}

\author{\speaker{Christopher Dilks}\\
        for the STAR Collaboration\\
        Pennsylvania State University / Duke University\\
        E-mail: \email{christopher.dilks@duke.edu}}

\abstract{ A significant sample of $\process$ events has been observed at STAR
in $\sqrt{s}=200$ GeV transversely polarized $pp$ collisions, where an isolated
$\pion$ is detected in the forward pseudorapidity range $2.65<\eta<3.9$ along
with the forward-going proton $p$, which scatters with a near-beam forward
pseudorapidity into Roman Pot detectors. The sum of the $\pion$ and the
scattered proton energies is consistent with the incident proton energy of 100
GeV, indicating that no further particles are produced in this direction. It is
postulated that the forward incident proton may have fluctuated into a $p+\pion$
system, with an angular momentum correlated with the initial proton spin.  The
backward-going proton interacts with the $p+\pion$ system, which then separates
such that the $\pion$ has a transverse momentum of ${\sim}2$ GeV/$c$ and the
proton has a transverse momentum of ${\sim}0.2$ GeV/$c$, while the backward
proton shatters into the remaining particles $X$.  Correlations between the
$\pion$ and scattered proton will be presented, along with single-spin
asymmetries which depend on the azimuthal angles of both the pion and the
proton.  This is the first time that spin asymmetries have been explored for
this process, and a model to explain their azimuthal dependence is needed.  }

\FullConference{XXVII International Workshop on Deep-Inelastic Scattering and Related Subjects - DIS2019\\
		8-12 April, 2019\\
		Torino, Italy}

\begin{document}

\section{Motivation} 

The transverse single-spin asymmetry, $A_N$, is an observable that
probes the spin structure of the proton. It is defined via 
\begin{equation}
A\left(\phi\right)=\frac
{d\sigma^{\uparrow}\left(\phi\right)-d\sigma^{\downarrow}\left(\phi\right)}
{d\sigma^{\uparrow}\left(\phi\right)+d\sigma^{\downarrow}\left(\phi\right)}
=A_N\cos\phi,
\label{eqAN}
\end{equation}
where $d\sigma^{\uparrow(\downarrow)}\left(\phi\right)$ is a differential cross
section, {\it e.g.}, for $\pion$ production, with azimuthal angle $\phi$, from a
spin-up(down) proton $p^{\uparrow(\downarrow)}$ scattering off an unpolarized
proton. The spin
asymmetry $A\left(\phi\right)$ is modulated by $\cos\phi$, and the amplitude is
denoted by $A_N$. If $\phi=0$ represents leftward $\pion$ production, then a
positive $A_N$ indicates spin-up(down) proton scattering favors producing
$\pion$s to the left(right).

$A_N$ for forward $\pion$s rises with Feynman-$x$ and is independent of center-of-mass
energy $\sqrt{s}$ \cite{xF1,xF2}; moreover, $A_N$ is systematically larger for
isolated $\pion$s than for those not as isolated \cite{heppelmannAN,mondalAN}.
Several models have been proposed to explain the origin of this large $A_N$
\cite{sivers1,sivers2,collins,twist3}, and although the most promising of these
involves a novel twist-3 fragmentation process \cite{twist3}, the origin of the
$\pion$-isolation dependence remains unclear.

\begin{figure}[t]
\centerline{\includegraphics[width=0.7\textwidth]{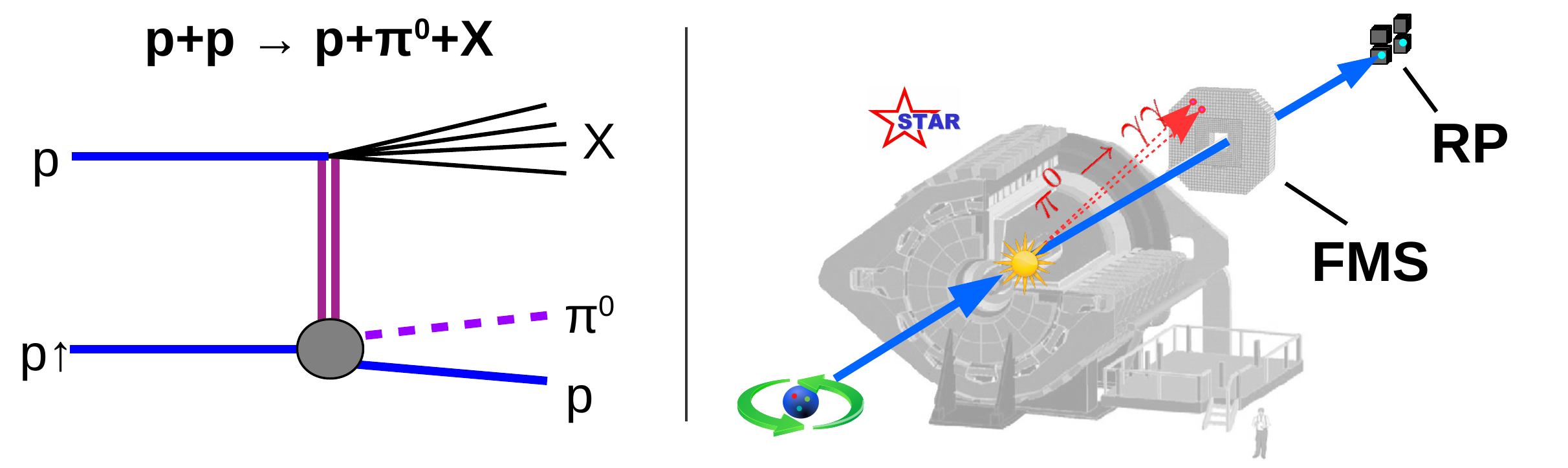}}
\caption{
Left: schematic of $\process$. Right: schematic of detectors, the Forward Meson
Spectrometer (FMS) for the $\pion\to\gamma\gamma$ (red dashed arrows) and the Roman
Pots (RP) for the proton (blue solid arrow).
}
\label{fig1}
\end{figure}

A possible channel for isolated $\pion$ production is the $\process$ process,
as shown schematically in the left panel of figure \ref{fig1}.  The forward
polarized proton $p^{\uparrow}$ scatters off the backward proton $p$; the
forward proton is deflected slightly with the production of a forward $\pion$,
while the backward proton fragments into remnants denoted by $X$. By energy
  conservation, the sum of the deflected proton and forward $\pion$ energies is
  equal to or less than the incident proton energy, while the observed $\pion$
  and proton transverse momentum sum should balance that of $X$. 

Further study is needed to understand the $\process$ underlying mechanism, and
especially its spin dependence. One possible model assumes the $p^{\uparrow}$
fluctuates into a $p+\pion$ state, with the $\pion$ in the proton periphery; if
the $\pion$ scatters off another proton such that the $p+\pion$ state separates,
then the $\pion$ could scatter with a moderate $p_T$, while the proton
recoils at near-beam rapidity. It is thought that the proton angular momentum
in the peripheral region is likely dominantly from orbital angular momentum,
rather than from parton spin \cite{weiss}; assuming the orbital angular momentum
of the peripheral $\pion$ correlates to the proton spin, measurements of spin
asymmetries in the $\process$ process could be sensitive to proton peripheral
angular momentum.

\section{Event Selection and Kinematics}

The $\process$ process has recently been observed at STAR in
transversely-polarized proton-proton scattering at $\sqrt{s}=200$ GeV during
the 2015 RHIC run. The $\pion$ is measured with the Forward Meson Spectrometer
(FMS), a lead-glass electromagnetic calorimeter subtending the forward region
$2.65<\eta<3.9$ \cite{aLL}, and the deflected proton with the Roman Pots (RP),
hodoscopic silicon-strip trackers downstream of the FMS, at near-beam
rapidity \cite{rp1,rp2}. The right panel of figure \ref{fig1} shows the
detectors, with overlaying $\pion\to\gamma\gamma$ and proton trajectories.

The $\pion$s were selected from each event's highest-energy photon pair, with a
transverse momentum $p_T$ above the trigger threshold and energy $E_1+E_2>12$
GeV. The invariant mass was constrained to the $\pion$ mass region and the
photons' energy imbalance to $\left|E_1-E_2\right|/\left(E_1+E_2\right)<0.8$.
The proton was required to be detected in at least 7 of the 8 available silicon
tracking planes, within geometric acceptance cuts, along with a veto on activity
in the RPs in the backwards beam direction. 

The selected events included a large contribution from accidental coincidences,
for example, two collisions occurring in a single proton bunch crossing, where
  one collision sent a $\pion$ to the FMS while the second one was elastic,
  sending a proton to the RPs. For many of these accidental coincidences, the
  sum of the $\pion$ and proton energies, $\esum:=\xh{E}+\xp{E}$, is greater
  than the 100 GeV incident proton energy, which would violate energy
  conservation had the proton and $\pion$ originated from the same collision.
  The Beam Beam Counters (BBC), scintillators in both the forward and backward
  directions subtending $2.1<|\eta|<5$, were used with cuts set to reduce the
  level of accidental coincidences while minimizing the loss of $\process$
  candidates. Moreover, evidence of hits in the backward BBC as well as in the
  central-rapidity Time Of Flight (TOF) detector was seen for all
  $\process$ events, indicating breakup of the backward-going proton.

\begin{figure}[t]
\centerline{\includegraphics[width=\textwidth]{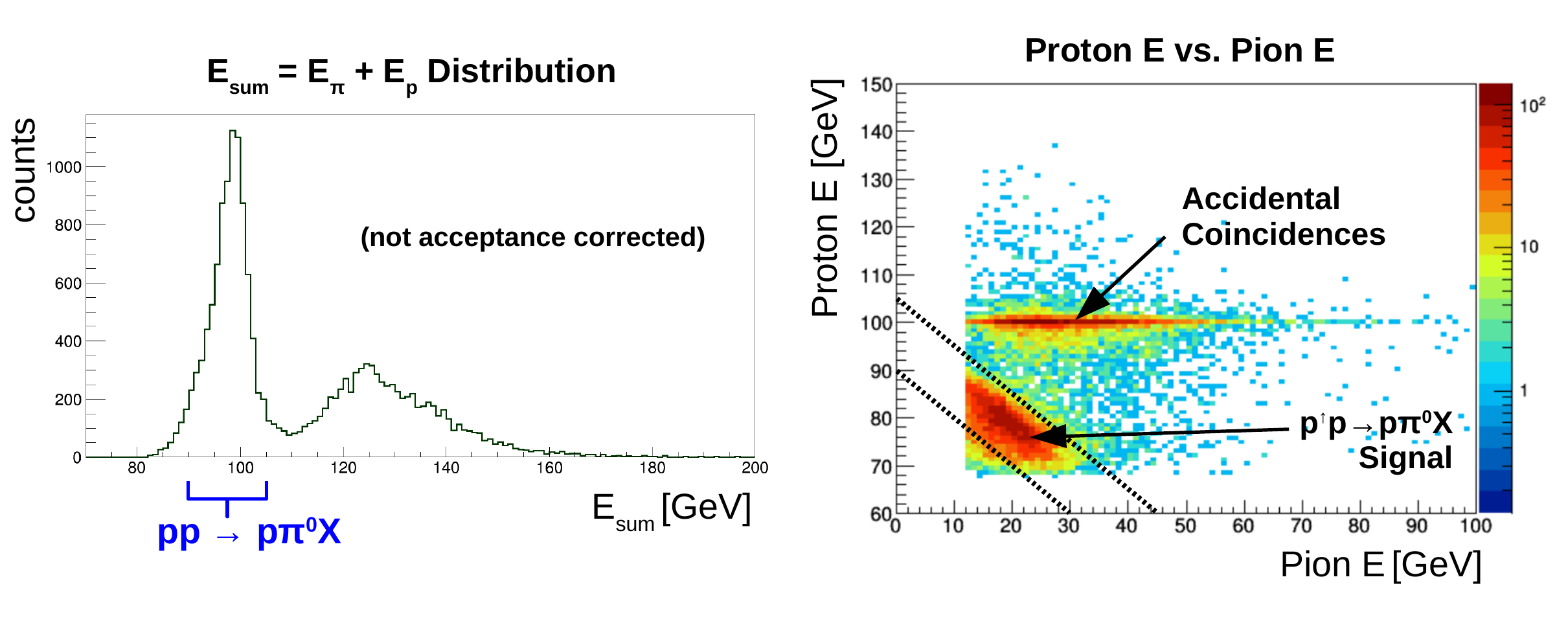}}
\caption{
Left: distribution of summed $\pion$ and proton energies, $\esum$, shown with
the $\process$ selection region. Right: proton energy on the vertical axis plotted
against $\pion$ energy; the region between the dashed lines is the $\process$
selection region.
}
\label{fig2}
\end{figure}

The left panel of figure \ref{fig2} shows a distribution of $\esum$, and the
right panel shows $\xp{E}$ plotted on the vertical axis versus $\xh{E}$ on the
horizontal.  The peak at $\esum=100$ GeV represents the $\process$ signal
region, since the incident proton has an energy of 100 GeV and, by energy
conservation, nothing else scattered in the forward direction; it corresponds to
the region between the dashed lines in the right panel.  The width of the 100
GeV $\esum$ peak is dominantly from the FMS energy resolution and an event
selection of $90<\esum<105$ GeV was used for asymmetry analysis event selection.

Since the RPs were designed to see elastic and diffractive-like events, the
$\xp{E}$ distribution has a large peak at $\xp{E}=100$ GeV, which manifests as a
band that spans the full $\xh{E}$ range.  These events along with any others
with $\esum$ above the $\process$ signal region are accidental coincidences, and
their $\esum$ distribution likely extends to low $\esum$ as the dominant source
of background under the $\process$ peak. The aforementioned BBC cut was tuned to
minimize the accidental coincidence background distribution and maximize the
$\process$ signal purity.

The resulting events have the following kinematics: the $\pion$ and proton
transverse momenta respectively span $1<p_{T,\pi}<4$ GeV/$c$ and
$0.1<p_{T,p}<0.45$ GeV/$c$, while their energies span $12<\xh{E}<35$ GeV and
$68<\xp{E}<90$ GeV.  For about $2/3$ of the events, the $\pion$ and proton are
observed back-to-back, with azimuthal angles $\phih$ and $\phip$ such that
$\delphi:=\phih-\phip\sim\pi$.  While the FMS spans the full $2\pi$ azimuth, the RP
silicon tracking planes are positioned above and below the beam, and
$\phip\sim0$ and $\phip\sim\pm\pi$, respectively left and right, are outside the
RP acceptance.

There is a further limit on $\phip$, since the RPs are positioned downstream of
a RHIC dipole magnet that bends the outgoing beam to the left.  This magnet is
tuned to bend beam-energy protons appropriately, so any scattered proton with
$\xp{E}\sim100$ GeV is likely to pass within the horizontal extent of the RPs.
The $\process$ events, however, have protons with $\xp{E}<90$ GeV, which are
bent more leftward than the 100 GeV protons.  Therefore the azimuthal acceptance
is biased toward rightward-scattered protons: $\pi/2<|\phip|<\pi$ for $90\%$ of
the events.  Despite this bias, it is still possible to analyze spin asymmetries
which depend on both $\phih$ and $\phip$; an upgraded RP system is required to
characterize $\process$ events with full proton azimuthal acceptance.

\section{Asymmetries}

Spin asymmetries of the $\process$ process can be modulated by two possible
azimuthal angles: $\phih$ and $\phip$.  In general, asymmetries and cross
sections can depend on the incident $p^{\uparrow}$ momentum vector $\vec{Z}$,
the observed $\pion$ and proton momentum vectors, respectively $\vec{\Pi}$ and
$\vec{P}$, and the $p^{\uparrow}$ spin pseudovector $\vec{S}$ with spin
projection $s=\pm\hbar/2$.  Physically allowed terms must be Lorentz invariant
and parity conserving, {\it i.e.} scalar, which can be formed by geometric
products of momenta and spin.  Asymmetry contributions must also depend on spin
$s$ and be invariant under rotations. For inclusive $\pion$ production, the
scalar $\left(\vec{Z}\times\vec{\Pi}\right)\cdot \vec{S}\propto s\cos{\phih}$
represents the $\pion$ transverse single-spin asymmetry $A_N$ of equation
\ref{eqAN}.

In $\process$, the additional proton momentum allows for the construction of
scalars which depend on both $\phip$ and $\phih$. Letting
$\vec{L}_{\pi}:=\vec{Z}\times\vec{\Pi}$ and $\vec{L}_p:=\vec{Z}\times \vec{P}$,
a possible scalar that satisfies the aforementioned requirements and depends on
both $\phip$ and $\phih$ is 
\begin{equation}
\left(\vec{L}_{\pi}\cdot\vec{L}_p\right)
\left(\vec{L}_p\cdot \vec{S}\right)
\propto s\modulation,
\label{eqAsym}
\end{equation}
which represents the transverse single-spin asymmetry of the $\pion$ within the
scattering plane of the observed proton. Letting $\asym$ denote the amplitude of
this modulation, $\left|\asym\right|$ is large when the
proton scatters left or right ($\phip\sim 0$ or $\pi$) and when the $\pion$ is
close to the proton scattering plane ($\delphi\sim0$ or $\pi$). Other possible
scalars were tested, but their measured asymmetries were consistent with zero.

\begin{figure}[t]
\centerline{\includegraphics[width=0.5\textwidth]{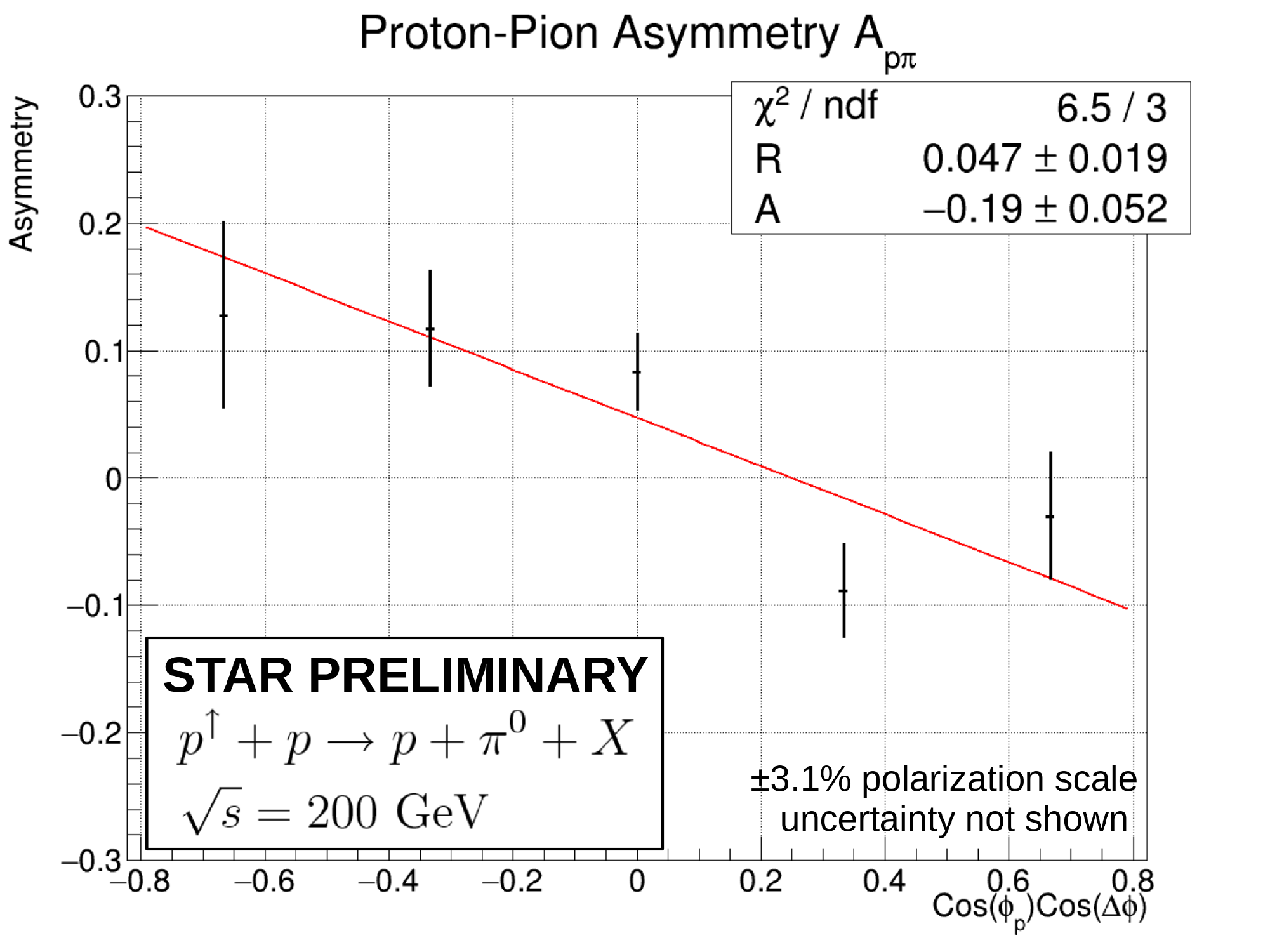}}
\caption{
Transverse single-spin asymmetry in bins of $\modulation$.  A linear fit is
included, with constant term $R$ and slope $A$, and the resulting fit values in
the upper right corner.
}
\label{fig4}
\end{figure}

Let $N^{\uparrow(\downarrow)}\left(\phih,\phip\right)$ denote the yield from
a spin-up(down) proton which scatters to a $\pion$ and proton with respective
azimuthal angles $\phih$ and $\phip$. With $P$ denoting the beam polarization, the
single-spin asymmetry was measured following equation \ref{eqAN} as
\begin{equation}
A\left(\phih,\phip\right)=\frac{1}{P}
\frac{N^{\uparrow}\left(\phih,\phip\right)-N^{\downarrow}\left(\phih,\phip\right)}
      {N^{\uparrow}\left(\phih,\phip\right)+N^{\downarrow}\left(\phih,\phip\right)}.
\label{eqAppi}
\end{equation}
Figure \ref{fig4} shows $A\left(\phih,\phip\right)$ in bins of
$\modulation$, including a linear fit with a slope that corresponds to the
amplitude of the $\modulation$ modulation, $\asym$, which evaluates to
$-19\%\pm5.2\%$. The fit's constant term $R$ is
included to account for possible nonzero relative luminosity which would
systematically shift all data points upward or downward across all $\modulation$
bins. The vertical error bars represent statistical uncertainty, and the
horizontal error bars are the combined propagated $\pion$ and proton position
uncertainties. The average beam polarization was $56.5\%$ and its uncertainty
propagates to a $3.1\%$ systematic uncertainty on the
asymmetry scale.

\begin{figure}[t]
\centerline{\includegraphics[width=\textwidth]{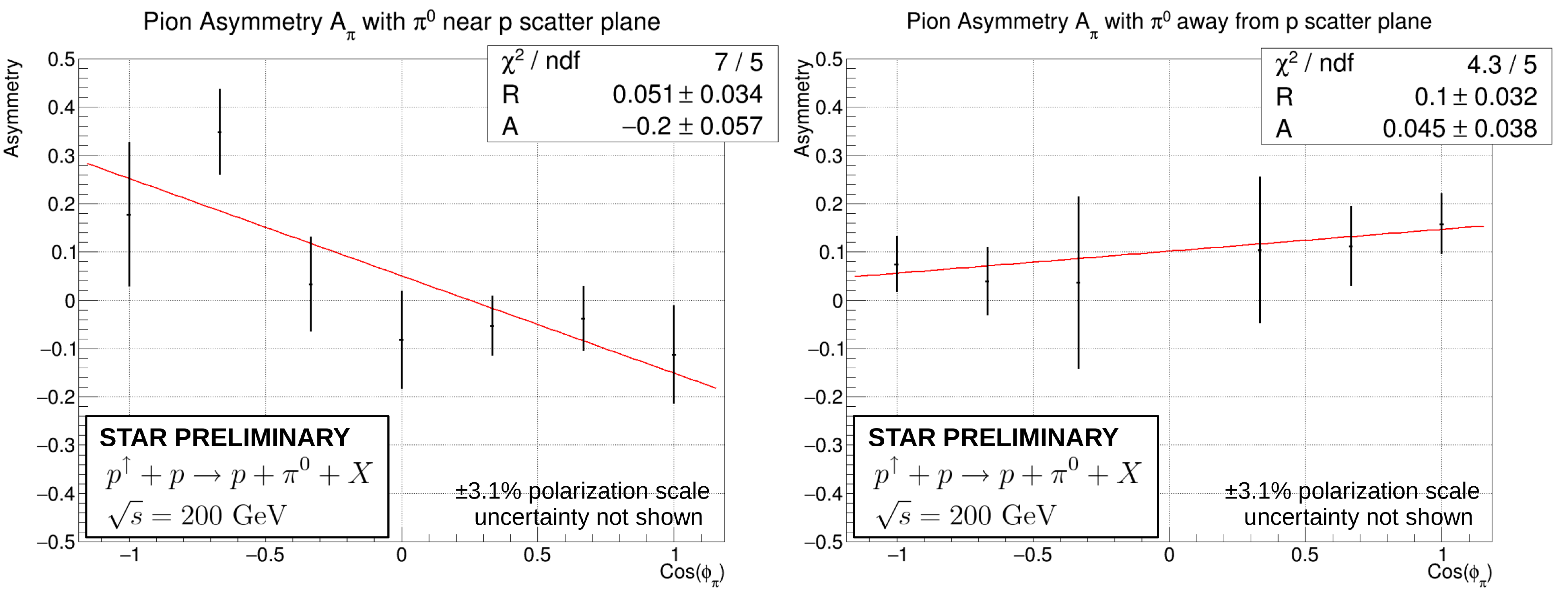}}
\caption{
Transverse single-spin asymmetry in bins of $\cos\phih$
for $\pion$s near the proton scattering plane (left) or away (right). A linear
fit is included in each.
}
\label{fig5}
\end{figure}

A complementary view of this asymmetry is shown in figure \ref{fig5}, where the
$\cos\phih$ modulation ($\pion$ $A_N$) is shown for $\pion$s which scatter near
the proton scattering plane (left panel), where $\delphi$ is within $\pi/6$
radians of $0$ or $\pm\pi$, compared to the case where $\pion$s scatter away
from the proton scatter plane (right panel), where
$\left|\delphi\pm\pi/2\right|<\pi/6$. When the $\pion$ scatters near the proton
scatter plane, it shows an asymmetry of $-20\%\pm5.7\%$, whereas when the
$\pion$ scatters out-of-plane, the asymmetry is nearly consistent with zero, at
$4.5\%\pm3.8\%$.

Projections of $\asym\modulation$ onto $\phih$, $\phip$, and $\delphi$ were used
to assess the impact of the limited $\phip$ acceptance; these are projections of
a 2-dimensional asymmetry to 1-dimensional asymmetries and can be cross-checked
with the corresponding 1-dimensional asymmetries in the data. Assuming the
nominal value of $\asym=-0.19$, projections of $\asym\modulation$ onto
1-dimensional asymmetries modulated by $\phih$, $\phip$, or $\delphi$ agree with
data only when the $\phip$ acceptance limitations are applied. While the
1-dimensional asymmetries are dependent on the $\phip$ acceptance limitations,
the 2-dimensional $\asym$ asymmetry is not and seems to most closely match the
data.  Several other possibilities were tested, such as the assumption that the
asymmetry is just a $\pion$ single-spin asymmetry, however their projections do
not agree with the data.

\section{Summary}

The $\process$ process has been observed at STAR, and a $-19\%\pm5.2\%$ asymmetry
of the $\pion$ in the scattering plane of the proton is observed, via the
modulation in equation \ref{eqAsym}. This effect may serve as a probe to the
orbital angular momentum of fluctuated $\pion$s in the proton periphery. As far
as we know, the spin-dependence of this process has otherwise not yet been
explored experimentally and a model is needed to understand it. Moreover, this
process should be studied in more detail experimentally, with better azimuthal
and kinematic coverage.

\end{document}